\documentclass[eqsecnum,twocolumn,aps,epsf]{revtex4}
\usepackage{graphicx}
\usepackage{color}

\newcommand{\LSCO}{La$_{2-x}$Sr$_x$CuO$_4$}
\newcommand{\LSCNO}{La$_{2-x}$Sr$_x$Cu$_{1-y}$Ni$_y$O$_4$}

\newcommand{\Tc}{$T_{\mathrm{c}}$}

\newcommand{\Cutwoplus}{Cu$^{2+}$}
\newcommand{\Nitwoplus}{Ni$^{2+}$}

\newcommand{\CCphT}{$(C_{\mathrm{tot}} - C_{\mathrm{ph}})/T$}

\newcommand{\Ctot}{$C_{\rm tot}$}

\newcommand{\gLT}{$\gamma _{\mathrm{LT}}$}
\newcommand{\gN}{$\gamma _{\mathrm{N}}$}

\newcommand{\peff}{$p_{\mathrm{eff}}$}

\newcommand{\y}{$y$}

\newcommand{\chiAF}{$\chi _{\mathrm{AF}}$}

\begin{document}
\draft

\title{Hole-trapping by Ni, Kondo effect and electronic phase diagram in non-superconducting Ni-substituted La$_{2-x}$Sr$_x$Cu$_{1-y}$Ni$_y$O$_4$}

\author{K. Suzuki}
\author{T. Adachi}
\thanks{Corresponding author: adachi@teion.apph.tohoku.ac.jp}
\author{Y. Tanabe}
\altaffiliation[Present address: ]{WPI-Advanced Institute of Materials Research, Tohoku University, 6-3 Aoba, Aramaki, Aoba-ku, Sendai 980-8579, Japan}
\author{Y. Koike}
\affiliation{Department of Applied Physics, Graduate School of Engineering, Tohoku University, 6-6-05 Aoba, Aramaki, Aoba-ku, Sendai 980-8579, Japan}

\author{T. Kawamata}
\altaffiliation[Present address: ]{Division of Material Science and JST, TRIP, Nagoya University, Furo-cho, Chikusa-ku, Nagoya 464-8602, Japan}
\affiliation{Advanced Meson Science Laboratory, Nishina Center for Accelerator-Based Science, RIKEN, 2-1 Hirosawa, Wako 351-0198, Japan}
\author{Risdiana}
\affiliation{Advanced Meson Science Laboratory, Nishina Center for Accelerator-Based Science, RIKEN, 2-1 Hirosawa, Wako 351-0198, Japan}
\affiliation{Department of Physics, Faculty of Mathematics and Natural Sciences, Padjadjaran University, Jl. Raya Bandung-Sumedang Km. 21 Jatinangor, Sumedang 45363, Indonesia}
\author{T. Suzuki}
\altaffiliation[Present address: ]{Graduate School of Arts and Sciences, International Christian University, 3-10-2 Osawa, Mitaka, Tokyo 181-8585, Japan}
\author{I. Watanabe}
\affiliation{Advanced Meson Science Laboratory, Nishina Center for Accelerator-Based Science, RIKEN, 2-1 Hirosawa, Wako 351-0198, Japan}

\date{\today}

\begin{abstract}

In order to investigate the electronic state in the normal state of high-{\Tc} cuprates in a wide range of temperature and hole-concentration, specific-heat, electrical-resistivity, magnetization and muon-spin-relaxation ($\mu$SR) measurements have been performed in non-superconducting Ni-substituted La$_{2-x}$Sr$_x$Cu$_{1-y}$Ni$_y$O$_4$ where the superconductivity is suppressed through the partial substitution of Ni for Cu without disturbing the Cu-spin correlation in the CuO$_2$ plane so much. 
In the underdoped regime, it has been found that there exist both weakly localized holes around Ni and itinerant holes at high temperatures.
With decreasing temperature, all holes tend to be localized, followed by the occurrence of variable-range hopping conduction at low temperatures.
Finally, in the ground state, it has been found that each {\Nitwoplus} ion traps a hole strongly and that a magnetically ordered state appears.
In the overdoped regime, on the other hand, it has been found that a Kondo-like state is formed around each {\Nitwoplus} spin at low temperatures.
In conclusion, the ground state of non-superconducting La$_{2-x}$Sr$_x$Cu$_{1-y}$Ni$_y$O$_4$ changes upon hole doping from a magnetically ordered state with the strong hole-trapping by {\Nitwoplus} to a metallic state with Kondo-like behavior due to {\Nitwoplus} spins, and the quantum phase transition is crossover-like due to the phase separation into short-range magnetically ordered and metallic regions.

\end{abstract}
\vspace*{2em}
\pacs{PACS numbers:74.25.Dw, 74.40.Kb, 74.72.Gh, 74.25.Bt}
\maketitle
\newpage

\section{Introduction}
In the history of the research of the high-$T_{\rm{c}}$ superconductivity (HTSC), studies of impurity-substitution effects have played one of central roles in the elucidation of the mechanism of HTSC. 
It is widely recognized in the hole-doped high-$T_{\rm{c}}$ cuprates that the suppression of the superconductivity through the substitution of magnetic impurities such as Ni for Cu is weaker than through the substitution of nonmagnetic impurities such as Zn for Cu,~\cite{Westerholt1989,Xiao1990} which is a trend opposite to that observed in conventional superconductors.~\cite{Tinkham}
Nuclear magnetic/quadrupole resonance measurements have revealed that Ni operates to weakly scatter holes, while Zn operates to strongly do.~\cite{Kitaoka1994} 
The so-called dynamical stripe correlations of spins and holes in the CuO$_2$ plane~\cite{tranquada} observed in La-based high-$T_{\rm{c}}$ cuprates tend to be pinned and stabilized by Zn more effectively than by Ni.~\cite{ada-niprb,ada-physc,Adachi2008} 
These contrasting behaviors suggest that Ni has weak influence on the electronic and magnetic states in the CuO$_2$ plane. 
In recent years, moreover, neutron scattering,~\cite{Matsuda2006,Hiraka2007} magnetization,~\cite{Machi2003} muon spin relaxation ($\mu$SR),~\cite{Tanabe2009} X-ray absorption fine structure (XAFS)~\cite{Hiraka2009} experiments and a theoretical work using the numerical exact diagonalization calculation~\cite{tsutsui} have suggested that a hole tends to be bound around a Ni impurity, leading to the decrease of the effective hole-concentration. 
That is, a Ni$^{2+}$ ion with the spin quantum number $S = 1$ tends to trap a hole, forming a Ni$^{2+}$ ion with a ligand hole, i.e., a so-called Zhang-Rice doublet state (the effective value of $S$ is 1/2) so as not to disturb the antiferromagnetic (AF) correlation between Cu$^{2+}$ spins with $S = 1/2$ so much. 
These suggest that Ni is no longer regarded as an usual magnetic impurity in the high-$T_{\rm{c}}$ cuprates.

The above new concept of hole-trapping by Ni may pin down the issue whether or not a quantum critical point (QCP) resides in the superconducting (SC) region on the phase diagram of high-$T_{\rm{c}}$ cuprates. 
It has been pointed out that the magnitude of the pseudo gap decreases with increasing hole-concentration per Cu, $p$, and seems to vanish at $p \sim 0.19$.~\cite{Tallon2001} 
Moreover, the electrical resistivity, $\rho$, has exhibited a $T$-linear behavior above $T_{\rm{c}}$ in a very limited region near the optimally doped regime, suggesting the existence of a quantum critical region.~\cite{Gurvitch1987,naqib,Ando2004} 
To investigate the quantum phase transition, the SC state concealing the normal ground state has to be removed. 
In the La-based cuprate La$_{2-x}$Sr$_x$CuO$_4$ (LSCO) where the superconductivity is suppressed by the application of high magnetic field, in fact, it has been suggested that the quantum critical region deduced from the $T$-linear behavior of $\rho$ seems not to converge at a singular point of $p$ in the ground state as in the case of a conventional quantum phase transition, but seems to spread out over a wide range of $p$ in the overdoped regime at low temperatures.~\cite{Cooper2009} 
In Zn-substituted La$_{2-x}$Sr$_x$Cu$_{1-y}$Zn$_y$O$_4$ (LSCZO) where the superconductivity is suppressed by Zn, on the other hand, $\mu$SR measurements~\cite{Risdiana2008} have revealed that the development of the Cu-spin correlation through the Zn substitution is weakened gradually with increasing $p$ in the overdoped regime and disappears around $x = 0.30$ where the superconductivity in LSCO disappears, suggesting that QCP in the viewpoint of the Cu-spin correlation~\cite{panago} is not located simply at $p \sim 0.19$ but is extended in the overdoped regime due to the presence of a phase separation into short-range magnetically ordered and metallic regions. 
Both high magnetic field and Zn suppressing the superconductivity, however, may affect the Cu-spin correlation which is probably indispensable to the appearance of HTSC.
Therefore, the exploration of QCP in \LSCNO (LSCNO) in which the superconductivity is suppressed by Ni attracts great interest.

In this paper, specific heat, $\rho$, magnetization and $\mu$SR measurements have been carried out to investigate the electronic state in the normal state of non-SC Ni-substituted LSCNO in a wide range of $x$ and $y$.~\cite{tanabe-NiJPSJ,suzuken} 
In particular, we have focused on the temperature-dependent change of the hole dynamics from the underdoped to overdoped regime in order to obtain a whole aspect of the electronic state in the normal state, because only separated results in limited ranges of $p$ and temperature have been reported so far.~\cite{Matsuda2006,Hiraka2007,Machi2003,Tanabe2009,Hiraka2009,tsutsui}

\section{Experimental}
Polycrystalline samples of LSCNO with $x = 0.08 - 0.30$ and $y = 0 - 0.10$ were prepared by the ordinary solid-state reaction method.~\cite{Adachi2008} 
All of the samples were checked by the powder X-ray diffraction to be of the single phase.
Specific-heat measurements were carried out by the thermal relaxation method in zero magnetic field and 9 T at low temperatures down to 0.4 K, using a commercial apparatus (PPMS, Quantum Design).
The $\rho$ measurements were performed by the standard four-probe method in zero field and 9 T at temperatures down to 0.4 K. 
Magnetization measurements were performed in a magnetic field of 1 T at temperatures down to 2 K, using a superconducting quantum interference device (SQUID) magnetometer (MPMS-XL5, Quantum Design).
In order to estimate the magnetic transition temperature, $T_{\rm N}$, zero-field $\mu$SR measurements were performed at temperatures down to 0.3 K at the Paul Scherrer Institute (PSI) in Switzerland and at the RIKEN-RAL Muon Facility at the Rutherford-Appleton Laboratory in the UK.

\section{Results}
\subsection{Specific heat}\label{sh}
\begin{figure}[tbp]
\begin{center}
\includegraphics[width=1.0\linewidth]{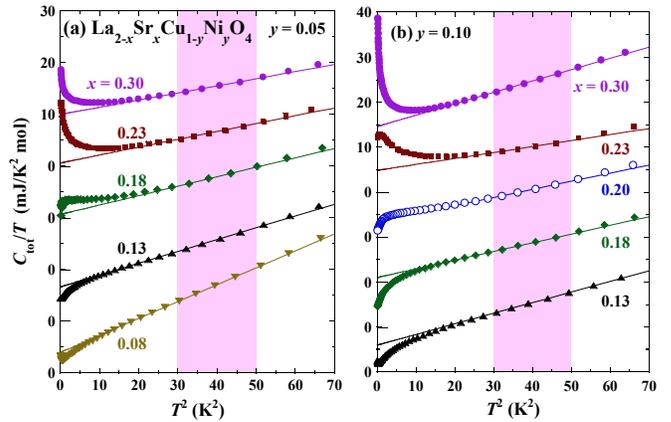}
\end{center}
\caption{(color online) Temperature dependence of the specific heat, $C_{\rm tot}$, for typical values of $x$ in La$_{2-x}$Sr$_x$Cu$_{1-y}$Ni$_y$O$_4$ with (a) $y = 0.05$ and (b) $ y = 0.10$ plotted as $C_{\rm tot}/T$ vs. $T^2$. The data are shifted top and bottom. Solid lines indicate the best-fit results obtained using Eq. (3.1). Shaded areas represent the fitting range of 30 K$^2 \le T^2 \le 50$ K$^2$ where $C_{\rm tot}/T$ is linear as a function of $T^2$.}  
\label{fig:fig1} 
\end{figure}

Figure 1 shows the temperature dependence of the specific heat, $C_{\rm tot}$, for LSCNO with $y=0.05$ and 0.10 plotted as $C_{\rm tot}/T$ vs. $T^2$. 
The data are shifted top and bottom. 
The specific heat in non-SC samples of LSCNO at low temperatures should be expressed as 

\begin{equation}
C_{\rm tot}(T)=\gamma T + \beta T^3. 
\label{eq:C/T}
\end{equation}
The first term represents the electronic specific heat, $C_{\rm el}$, and $\gamma$ is the electronic specific-heat coefficient proportional to the density of states (DOS) of quasiparticles at the Fermi level, described in the Fermi-liquid state. 
The second term represents the phonon specific heat, $C_{\rm ph}$, assuming the Debye model. 
In Fig. 1, every $C_{\rm tot}/T$ shows a $T^2$-dependence in a range of 30 K$^2 \le T^2 \le$ 50 K$^2$, indicating that every {\Ctot} is well expressed by Eq. (3.1). 
For $T^2 \le 30$ K$^2$, on the other hand, it is found that $C_{\rm tot}/T$ tends to deviate upward or downward from the $T^2$-dependence, suggesting an anomalous change of $C_{\rm el}$ at low temperatures. 

\begin{figure}[tbp]
\begin{center}
\includegraphics[width=0.8\linewidth]{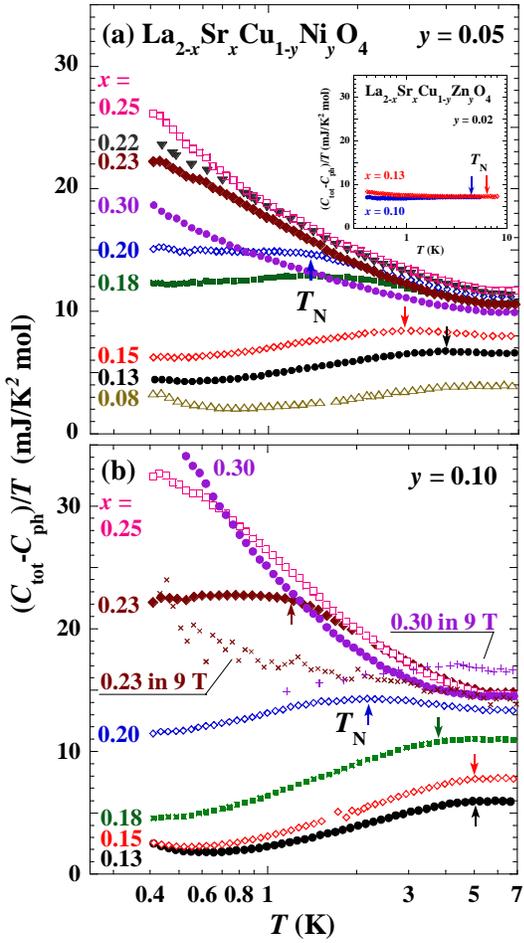}
\end{center}
\caption{(color online) Temperature dependence of $(C_{\rm tot}-C_{\mathrm{ph}})/T$ for La$_{2-x}$Sr$_x$Cu$_{1-y}$Ni$_y$O$_4$ with (a) $y = 0.05$ and (b) $y = 0.10$~\cite{suzuken} in zero field. Here, $C_{\mathrm{ph}}$ is the phonon specific heat  estimated from the best-fit results obtained using Eq. (\ref{eq:C/T}) in the temperature range of 30 K$^2 \le T^2 \le 50$ K$^2$. Data of $x=0.23, 0.30$ and $y=0.10$ in a magnetic field of 9 T are also indicated by crosses. The inset shows the temperature dependence of $(C_{\rm tot}-C_{\mathrm{ph}})/T$ in Zn-substituted La$_{2-x}$Sr$_x$Cu$_{1-y}$Zn$_y$O$_4$ with $x = 0.10$, 0.13 and $y = 0.02$. Arrows indicate the magnetic transition temperature, $T_{\mathrm{N}}$, estimated from $\mu$SR measurements.~\cite{Tanabe2009,ada-prb}}  
\label{fig:fig2} 
\end{figure}

After removing the phonon contribution from the total specific heat to investigate the anomalous deviation at low temperatures in detail, the temperature dependence of $(C_{\rm tot}-C_{\rm ph})/T$ for LSCNO with $y$ = 0.05 and 0.10 is shown in Figs. 2(a) and (b), respectively. 
Here, $C_{\rm ph}$ was estimated from the fit of the data in the temperature range of 30 K$^2 \le T^2 \le$ 50 K$^2$ in Fig. 1 to Eq. (3.1). 
For all the samples, $(C_{\rm tot}-C_{\rm ph})/T$ is found to be almost constant above 5 K, while it significantly varies depending on $x$ below 5 K. 
Here, we define the effective hole-concentration per Cu, \peff, in a sample as $p_{\rm{eff}} = x - y$, which is reasonable in the case that a \Nitwoplus ion traps a hole.
For low-{\peff} samples of $(x,y)=(0.08-0.13,0.05)$ and $(0.13-0.18,0.10)$, $(C_{\rm tot}-C_{\rm ph})/T$ decreases with decreasing temperature below 5 K, while it increases below 5 K for high-{\peff} samples of $(x,y)=(0.20-0.30, 0.05)$ and $(0.25-0.30, 0.10)$. 
For intermediate-{\peff} samples of $(x,y)=(0.15-0.18, 0.05)$ and $(0.20-0.23, 0.10)$, $(C_{\rm tot}-C_{\rm ph})/T$ once increases with decreasing temperature and then decreases below 5 K. 
Therefore, the behavior of $(C_{\rm tot}-C_{\rm ph})/T$ at low temperatures are summarized to exhibit a gradual crossover from the decrease to increase with increasing $x$. 
It is noted that a small upturn of $(C_{\rm tot}-C_{\rm ph})/T$ around the lowest temperature observed for $(x,y)=(0.08-0.13, 0.05)$ and $(0.13-0.15, 0.10)$ is probably due to the so-called Schottky anomaly often observed in high-$T_{\rm c}$ cuprates.~\cite{nohara} 

\begin{figure}[tbp]
\begin{center}
\includegraphics[width=1.0\linewidth]{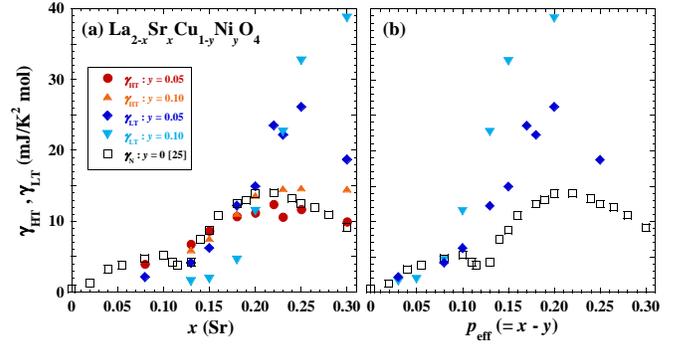}
\end{center}
\caption{(color online) (a) Sr-concentration $x$ dependence and (b) effective hole-concentration, $p_{\mathrm{eff}}$ ($= x - y$), dependence of $\gamma _{\mathrm{HT}}$ and $\gamma _{\mathrm{LT}}$ in La$_{2-x}$Sr$_x$Cu$_{1-y}$Ni$_y$O$_4$ with $y = 0.05$ and $y = 0.10$. Normal-state values of $\gamma$, $\gamma _{\mathrm{N}}$, in Ni-free {\LSCO} obtained by Momono and Ido~\cite{momono} are also plotted for comparison.}  
\label{fig:fig3} 
\end{figure}

To make quantitative discussion about $C_{\rm el}$, values of $(C_{\rm tot}-C_{\rm ph})/T$ before and after the change around 5 K are defined as $\gamma_{\rm HT}$ and $\gamma_{\rm LT}$, respectively, and plotted in Fig. 3(a). 
Concretely, $\gamma_{\rm HT}$ is defined as the averaged value of $(C_{\rm tot}-C_{\rm ph})/T$ at $5-7$ K for all the samples. 
The $\gamma_{\rm LT}$ for $(x,y)=(0.08-0.13, 0.05)$ and $(0.13-0.15, 0.10)$ exhibiting the Schottky anomaly is defined as the minimum value at low temperatures below 1 K and $\gamma_{\rm LT}$ for the other samples of $(x,y)=(0.15-0.30, 0.05)$ and $(0.18-0.30, 0.10)$ is defined as the value of {\CCphT} at the lowest temperature of 0.4 K. 
Normal-state values of $\gamma$, $\gamma_{\rm N}$, in Ni-free LSCO obtained by Momono and Ido~\cite{momono} is also plotted for comparison. 
It is found that $\gamma_{\rm HT}$ is in approximate agreement with $\gamma_{\rm N}$ at each $x$. 
This indicates that values of $\gamma$ are dependent on the Sr-concentration $x$ irrespective of the Ni-concentration $y$ at high temperatures above 5 K. 
Therefore, in comparison with the other experimental results suggesting the occurrence of hole-trapping by Ni at high temperatures,~\cite{Matsuda2006,Hiraka2007,Hiraka2009} the present results indicate that the complete hole-trapping by Ni does not take place at high temperatures above 5~K. 

On the other hand, $\gamma_{\rm LT}$ appears to disagree with $\gamma_{\rm N}$ at each $x$. 
Figure 3(b) shows the dependence on {\peff} of $\gamma_{\rm LT}$ and $\gamma_{\rm N}$. 
It is found that $\gamma_{\rm LT}$ is in approximate agreement with $\gamma_{\rm N}$ for low-{\peff} samples of $(x,y)=(0.08-0.15, 0.05)$ and $(0.13-0.18, 0.10)$, suggesting that the strong hole-trapping by {\Nitwoplus} takes place resulting in the effective decrease of the hole-concentration in these samples at low temperatures below 5~K. 

It is possible that the decrease in $(C_{\rm tot}-C_{\rm ph})/T$ with decreasing temperature at low temperatures below 5 K is simply caused by the reduction of DOS at the Fermi level accompanied by the formation of a magnetic order, not accompanied by the strong hole-trapping by Ni$^{2+}$. 
To check the possibility, the temperature dependence of the specific heat was measured for Zn-substituted LSCZO with $x=0.10, 0.13$ and $y=0.02$, in which a stripe-like magnetic order is formed at low temperatures below $T_{\rm N}$ estimated from $\mu$SR measurements,~\cite{ada-prb} as shown in the inset of Fig. 2(a). 
It is found that no significant decrease in $(C_{\rm tot}-C_{\rm ph})/T$ is observed below $T_{\rm N}$, suggesting that the observed decrease in $(C_{\rm tot}-C_{\rm ph})/T$ with decreasing temperature at low temperatures below 5 K for LSCNO is irrelevant to the magnetically induced reduction of DOS. 

\begin{figure}[tbp]
\begin{center}
\includegraphics[width=0.9\linewidth]{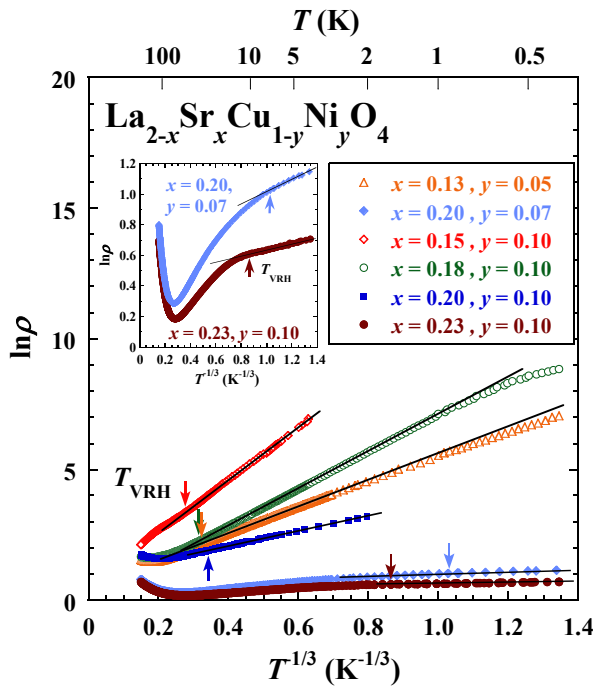}
\end{center}
\caption{(color online) Temperature dependence of the electrical resistivity, $\rho$, plotted as $\ln \rho$ vs. $T^{-1/3}$ for La$_{2-x}$Sr$_x$Cu$_{1-y}$Ni$_y$O$_4$ with $y = 0.05$, 0.07 and 0.10. The inset shows the data of $x = 0.20$, $y = 0.07$ and $x = 0.23$, $y = 0.10$ plotted on a different scale. Arrows indicate $T_{\mathrm{VRH}}$ below which $\ln \rho$ changes linearly to $T^{-1/3}$.}  
\label{fig:fig4} 
\end{figure}

For high-{\peff} samples of $(x,y)=(0.18-0.30, 0.05)$ and $(0.20-0.30, 0.10)$, $\gamma_{\rm LT}$ is apparently larger than $\gamma_{\rm N}$ owing to the enhancement of $(C_{\rm tot}-C_{\rm ph})/T$ at low temperatures below 5~K. 
The enhancement of $(C_{\rm tot}-C_{\rm ph})/T$ is reminiscent of the Kondo effect.~\cite{Kondo} 
As shown in Fig. 2(b), in fact, the enhancement of $(C_{\rm tot}-C_{\rm ph})/T$ at low temperatures for $(x,y)=(0.23, 0.10)$ and $(0.30, 0.10)$ is suppressed by the application of a magnetic field of 9 T, which is also a typical behavior in the Kondo state.
Therefore, high-{\peff} samples with Ni impurities are interpreted as being in a Kondo-like state at low temperatures below 5 K.
As for intermediate-{\peff} samples of $(x,y)=(0.15-0.18, 0.05)$ and $(0.20-0.23, 0.10)$, the enhancement of $(C_{\rm tot}-C_{\rm ph})/T$ with decreasing temperature is followed by the decrease below $\sim 1$ K. This is regarded as a crossover of the electronic state from the Kondo-like state to the strong hole-trapping state with decreasing temperature.

\subsection{Electrical resistivity}
Figure 4 shows the temperature dependence of $\rho$ plotted as ln$\rho$ vs. $T^{-1/3}$ for $(x,y)=(0.13, 0.05), (0.20, 0.07)$ and $(0.15-0.23, 0.10)$. 
Linear behaviors are observed at low temperatures below $T_{\rm VRH}$ shown by arrows for all the samples, suggesting the occurrence of two-dimensional (2D) variable-range hopping (VRH) conduction. 
As $\rho$ exhibits a metallic behavior at high temperatures for $(x,y)=(0.13, 0.05), (0.20, 0.07)$ and $(0.18-0.23, 0.10)$, it is suggested for low-$p_{\rm eff}$ samples that holes tend to be localized gradually with decreasing temperature and turn into a strongly localized state exhibiting VRH conduction below $T_{\rm VRH}$. 

\begin{figure}[tbp]
\begin{center}
\includegraphics[width=0.8\linewidth]{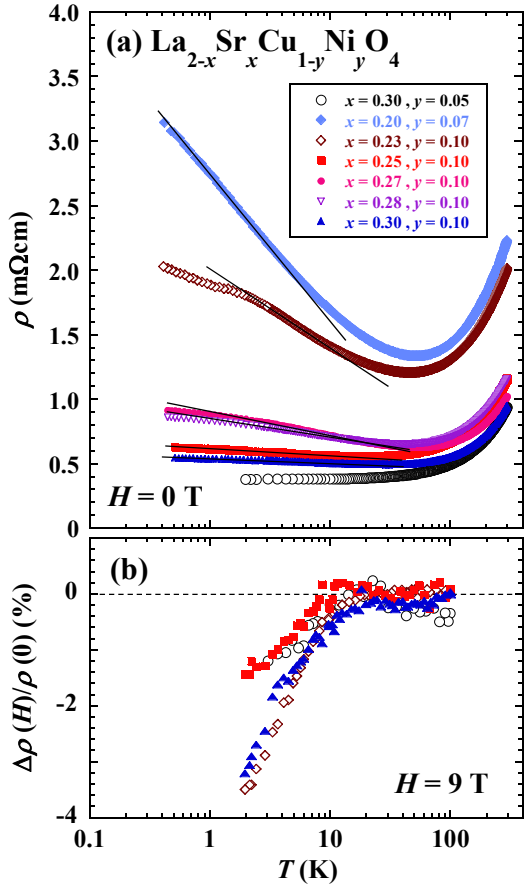}
\end{center}
\caption{(color online) Temperature dependence of (a) the electrical resistivity, $\rho$, in zero field and (b) the magnetoresistance, $\Delta \rho(H)/\rho(0) \equiv \{ \rho (H, T)-\rho (0,T)\}/\rho (0,T)$, in a magnetic field of 9 T in La$_{2-x}$Sr$_x$Cu$_{1-y}$Ni$_y$O$_4$ with $y = 0.05, 0.07$ and $0.10$. }  
\label{fig:fig5} 
\end{figure}

Figure 5(a) shows the temperature dependence of $\rho$ for $(x,y)=(0.30, 0.05), (0.20,0.07)$ and $(0.23-0.30, 0.10)$. 
It is found that the samples except for $(x,y)=(0.30, 0.05)$ exhibit logarithmic temperature-dependence below $5-10$ K and following saturation below $1-2$ K. 
As shown in Fig. 5(b), moreover, negative magnetoresistance is observed at low temperatures where the logarithmic temperature-dependence of $\rho$ is observed. 
These are characteristic of a Kondo state and/or a 2D Anderson-localized state. 
In these samples, in fact, a behavior characteristic of a Kondo state is observed also in the specific heat measurements, as described in Sec. \ref{sh}. 
Therefore, the logarithmic temperature-dependence of $\rho$ in zero field is interpreted as being due to the occurrence of the Kondo-like state, namely, due to screening of localized {\Nitwoplus} spins by mobile holes. 

For $(x,y)=(0.23, 0.10)$, it is found that the logarithmic increase in $\rho$ with decreasing temperature tends to be saturated around $1-2$ K and then increases again below 1 K. 
The saturated behavior of $\rho$ may be regarded as the one toward the unitary limit of the Kondo state. 
To understand the increase below 1 K, however, another mechanism is required, as discussed later.
It is noted for $(x,y)=(0.30, 0.05)$ that no logarithmic increase in $\rho$ is observed at low temperatures, although $(C_{\rm tot}-C_{\rm ph})/T$ increases with decreasing temperature below 5 K as shown in Fig. 2(a). 
Since negative magnetoresistance is observed as shown in Fig. 5(b), it is inferred that a Kondo-like state is realized at low temperatures for $(x,y)=(0.30, 0.05)$ as well.

\subsection{Magnetization}
\begin{figure}[tbp]
\begin{center}
\includegraphics[width=1.0\linewidth]{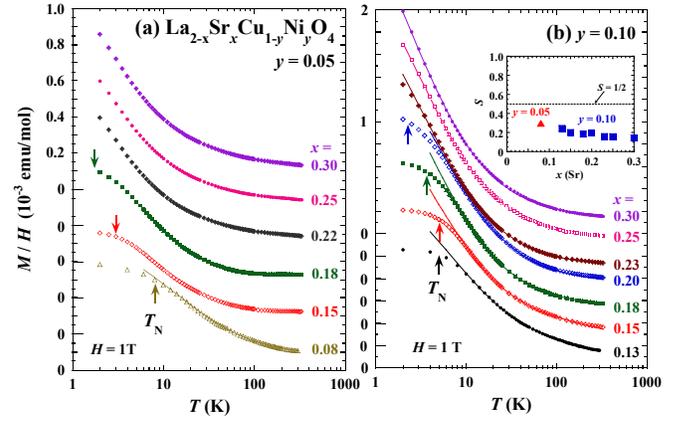}
\end{center}
\caption{(color online) Temperature dependence of the magnetization divided by the magnetic field, $M/H$, in a magnetic field of 1 T for La$_{2-x}$Sr$_x$Cu$_{1-y}$Ni$_y$O$_4$ with (a) $y = 0.05$ and (b) $y = 0.10$. The data are shifted top and bottom. Solid lines indicate the best-fit results obtained using Eq. (3.2) at high temperatures far from the magnetic transition temperature, $T_{\mathrm{N}}$, estimated from $\mu$SR measurements~\cite{Tanabe2009} shown by arrows. The inset shows the Sr-concentration $x$ dependence of the spin quantum number, $S$, estimated from the best-fit results.}  
\label{fig:fig6} 
\end{figure}

Figure 6 shows the temperature dependence of the magnetization divided by the magnetic field, $M/H$, for LSCNO with $y=0.05$ and 0.10.
The data are shifted top and bottom. 
The characteristics are as follows.
(i) All the samples exhibit Curie-like behavior more or less. 
(ii) For $(x,y)=(0.08-0.18, 0.05)$ and $(0.13-0.20, 0.10)$, downward deviation of $M/H$ from the Curie-like behavior is observed at low temperatures below $\sim 10$ K. 
(iii) The $M/H$ due to the so-called 2D-AF correlation between {\Cutwoplus} spins,~\cite{nakano} $\chi_{\rm AF}$, is invisible for all the samples, because probably the variation of the Curie-like $M/H$ with temperature is much larger than that of {\chiAF} in the measured temperature-range.

The temperature dependence of $M/H$ is fitted using the following equation; 
\begin{equation}
M/H = \chi_0 + \chi_{\rm AF} + C/(T-\theta).
\label{eq:chi}
\end{equation}
The first term represents a temperature-independent one originating from the van Vleck paramagnetism, ion-core diamagnetism, etc. 
The third term is a Curie-Weiss one due to {\Nitwoplus} spins. 
The $\theta$ is the Weiss temperature and $C$ the Curie constant given by
\begin{equation}
C = N g^2 S(S+1) \mu_{\rm B}^2 / 3 k_{\rm B}.
\label{eq:c}
\end{equation}
Here, $N$ is the number of {\Nitwoplus} spins, $g$ the g-factor, $\mu_{\rm B}$ the Bohr magneton and $k_{\rm B}$ the Boltzmann constant. 
Early measurements of $M/H$ in LSCNO by Xiao {\it et al}.~\cite{Xiao1990} have suggested that the Curie-like behavior is caused by the substituted {\Nitwoplus} spins.

For underdoped and highly Ni-substituted samples of LSCNO, it has been reported that $\chi_{\rm AF}$ exhibits a broad maximum above room temperature and monotonically decreases with decreasing temperature below room temperature,~\cite{nakano} allowing us to express $\chi_{\rm AF}$ being linearly dependent on temperature. 
For optimally doped and overdoped samples of LSCNO, on the other hand, the broad maximum in $\chi_{\rm AF}$ has been observed below room temperature.~\cite{nakano}
In fact, the fitting of the data of $(x,y) = (0.15-0.30,0.05)$ to Eq. (\ref{eq:chi}) in which $\chi_{\rm AF}$ is linearly dependent on temperature has failed.
On the other hand, the data of $(x,y)=(0.08, 0.05)$ and $(0.13-0.30, 0.10)$ have been fitted to Eq. (\ref{eq:chi}) at high temperatures above $T_{\rm N}$,  as shown in Fig. 6. 
Estimated values of $S$ using Eq. (\ref{eq:c}) are plotted in the inset of Fig. 6. 
It is found that the values of $S$ are between 0.1 and 0.3 and much smaller than the expected value of $S=1$ in the Ni$^{2+}$ state. 
This suggests that holes tend to be localized around {\Nitwoplus} forming Zhang-Rice doublet states even at high temperatures for comparatively low-{\peff} samples, so that each {\Nitwoplus} spin is antiferromagnetically coupled with a hole-spin, resulting in the decrease in $S$ of {\Nitwoplus} spins. 
It is noted that the smaller values of $S$ than 1/2 expected from the AF coupling between a {\Nitwoplus} spin with $S=1$ and a hole-spin with $S=1/2$ may be due to the mixing of Ni 3$d$ and O 2$p$ orbitals. 

\begin{figure}[tbp]
\begin{center}
\includegraphics[width=1.0\linewidth]{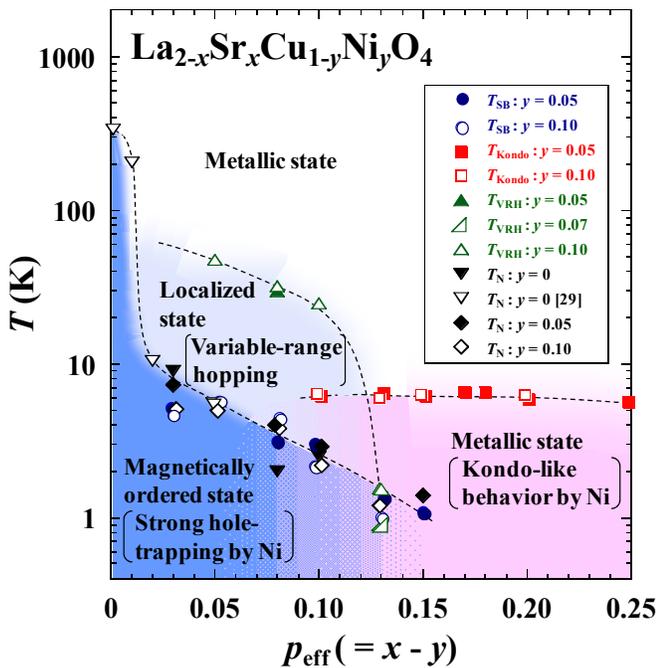}
\end{center}
\caption{(color online) Electronic phase diagram of non-superconducting Ni-substituted La$_{2-x}$Sr$_x$Cu$_{1-y}$Ni$_y$O$_4$.~\cite{suzuken} $T_{\mathrm{SB}}$ is defined as the temperature below which $(C_{\rm tot}-C_{\mathrm{ph}})/T$ decreases with decreasing temperature. $T_{\mathrm{Kondo}}$ is the onset temperature of the Kondo-like state where $(C_{\rm tot}-C_{\mathrm{ph}})/T$ increases with decreasing temperature. $T_{\mathrm{VRH}}$ is defined as the onset temperature of the variable-range hopping conduction. The magnetic transition temperature, $T_{\mathrm{N}}$, estimated from $\mu$SR measurements~\cite{Tanabe2009} is also plotted, together with the reported data of $y=0$.~\cite{Budnick} Dashed lines are to guide the reader's eye. }  
\label{fig:fig7} 
\end{figure}

As to the downward deviation of $M/H$ from the Curie-like behavior at low temperatures for comparatively low-{\peff} samples, the onset temperature of the deviation of $M/H$ is found to be in rough agreement with $ T_{\rm N}$ estimated from $\mu$SR measurements.~\cite{Tanabe2009}
This suggests that the deviation of $M/H$ is due to the development of the magnetic correlation between {\Nitwoplus} and {\Cutwoplus} spins toward the formation of a magnetic order.

\section{Discussion}
\subsection{Phase diagram of non-SC Ni-substituted La$_{2-x}$Sr$_x$Cu$_{1-y}$Ni$_y$O$_4$}
Figure 7 displays the phase diagram of LSCNO, in which the temperatures where $(C_{\rm tot}-C_{\rm ph})/T$ shown in Fig. 2 starts to decrease and increase with decreasing temperature, $T_{\rm SB}$ and $T_{\rm Kondo}$, respectively, the temperature below which $\rho$ shown in Fig. 4 exhibits VRH conduction, $T_{\rm VRH}$, are plotted as a function of $p_{\rm eff}$. 
For high-{\peff} samples, it is unclear whether or not holes tend to be localized around {\Nitwoplus} at high temperatures, so that {\peff} may be meaningless. 
In the Kondo-like state at low temperatures, on the other hand, holes are localized in the ground state so as to screen {\Nitwoplus} spins, so that $p_{\rm eff}$ is meaningful at low temperatures.

In Fig. 7, $T_{\rm N}$ estimated from $\mu$SR measurements~\cite{Tanabe2009} is also plotted, together with the reported data of $y=0$.~\cite{Budnick}
For $p_{\rm eff} < 0.10$, a long-range magnetic order is formed in the ground state, confirmed by the muon-spin precession. 
For $p_{\rm eff} \ge 0.10$, on the other hand, both the presence of fast depolarization of muon spins and the absence of muon-spin precession at low temperatures suggest the formation of a short-range magnetic order, which is discussed later. 
It is found that $T_{\rm SB}$ is in good agreement with $T_{\rm N}$, suggesting that the decrease in the electronic specific heat at low temperatures is deeply related to the formation of the magnetic order. 
That is, supposed that a hole is strongly bound by a {\Nitwoplus} ion, the effective value of $S$ of the Ni$^{2+}$ ion with a ligand hole is regarded as being roughly the same as that of Cu$^{2+}$ spins, tending to the recovery of the $S=1/2$ network of {\Cutwoplus} spins.~\cite{tsutsui} Moreover, the strong hole-trapping by a {\Nitwoplus} ion is expected to reduce the frustration effect between {\Cutwoplus} spins.
Therefore, it is guessed that each {\Nitwoplus} ion traps a hole strongly so as not to disturb the AF correlation in the network of Cu$^{2+}$ spins.

\subsection{Temperature dependence of the electronic state in the underdoped regime}
\begin{figure}[tbp]
\begin{center}
\includegraphics[width=1.0\linewidth]{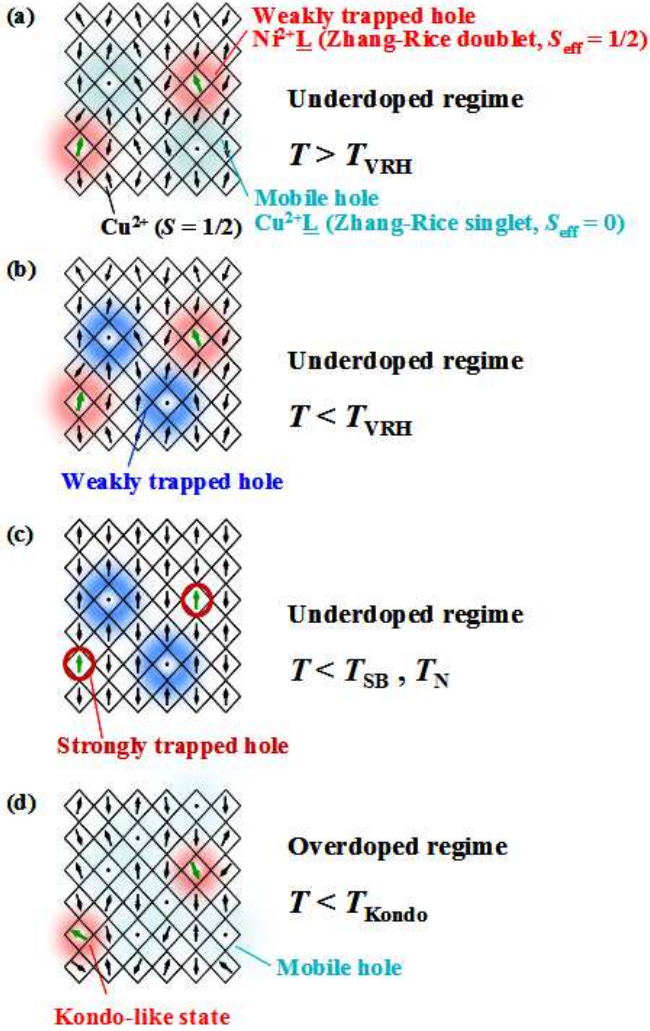}
\end{center}
\caption{(color online) Schematic drawing of the CuO$_2$ plane (a) in the underdoped regime at high temperatures of $T > T_{\mathrm{VRH}}$ where both weakly localized holes around {\Nitwoplus} and itinerant holes exist, (b) in the underdoped regime at $T < T_{\mathrm{VRH}}$ where localized holes exhibit variable-range hopping conduction, (c) in the underdoped regime at low temperatures below $T_{\mathrm{N}} \sim T_{\mathrm{SB}}$ where holes are strongly bound by {\Nitwoplus}, (d) in the overdoped regime at low temperatures below $T_{\rm Kondo}$ where screening of localized {\Nitwoplus} spins by conducting holes takes place forming a Kondo-like state.}  
\label{fig:fig8} 
\end{figure}

Here, we discuss the temperature-dependent change of the electronic state in the underdoped regime, as schematically shown in Fig. 8. 
In the case of $x>y$, that is, $p$ is larger than the Ni-concentration, {\y}, there exist two kinds of hole in the CuO$_2$ plane; one is weakly localized around {\Nitwoplus} forming a Zhang-Rice doublet state and the other is mobile forming a Zhang-Rice singlet state with {\Cutwoplus} spins at room temperature, as shown in Fig. 8(a). 
It has been suggested from XAFS experiments by Hiraka {\it et al}.~\cite{Hiraka2009} that holes tend to be localized around Ni even at room temperature. 
The present result of $M/H$ that the value of $S$ of {\Nitwoplus} spins is much smaller than 1 suggests the localization of holes around {\Nitwoplus} at high temperatures.
On the other hand, relatively good conduction occurs at high temperatures, judged from $\rho$ shown in Fig. 4. 
These results naturally draw the conclusion that the two kinds of hole, namely, weakly localized holes around {\Nitwoplus} forming a Zhang-Rice doublet state and mobile holes forming a Zhang-Rice singlet state with {\Cutwoplus} spins reside in the CuO$_2$ plane at high temperatures.

As shown in Fig. 8(b), mobile holes moving around in the CuO$_2$ plane at high temperatures tend to be localized below $T_{\rm VRH}$, leading to the occurrence of VRH conduction. 
On the other hand, holes around {\Nitwoplus} are maintained in its original form, supported by the constant value of $S$ in the moderate range of temperature as shown in Fig. 6 and by the good coincidence of $\gamma_{\rm HT}$ with $\gamma_{\rm N}$ at each $x$. 

With further decreasing temperature, each {\Nitwoplus} ion strongly traps a hole, resulting in the formation of the magnetic order of {\Cutwoplus} spins and in the decrease in $(C_{\rm tot}-C_{\rm ph})/T$ at low temperatures below $T_{\rm N} \sim T_{\rm SB}$, as shown in Fig. 8(c). 
Below $T_{\rm N}$, moreover, the suppression of $M/H$ is observed due to the participation of {\Nitwoplus} spins in the AF order of {\Cutwoplus} spins. 
No significant anomaly is detected in $\rho$ around $T_{\rm SB}$, implying that the state of holes except for ones around {\Nitwoplus} does not change so much.

\subsection{Temperature dependence of the electronic state in the overdoped regime}
In the overdoped regime, $\rho$ exhibits a metallic behavior at high temperatures.
At low temperatures below $T_{\rm Kondo}$, on the other hand, Kondo-like behavior is observed in the specific heat and $\rho$ measurements. 
In this case, mobile holes are scattered by {\Nitwoplus} spins, leading to the screening of the {\Nitwoplus} spins toward the ground state, as shown in Fig. 8(d). 
It attracts interest that the ground state of the Kondo-like state due to Ni$^{2+}$ spins in the overdoped regime is more or less analogous to the hole-trapping state due to Ni$^{2+}$ in the underdoped regime. 
Very recently, a logarithmic increase in $\rho$ and negative magnetoresistance have been observed also in the overdoped Bi-based cuprate in which Fe with a large magnetic moment is partially substituted for Cu, \cite{Wakimoto} suggesting the occurrence of the Kondo effect.
Here, it should be stressed that the Kondo state generally appears in metallic samples. 
Therefore, the present result is one of significant evidences for the occurrence of a Fermi-liquid state in the overdoped regime.

\subsection{Change of the ground state upon hole doping}
In the ground state of the underdoped regime, each {\Nitwoplus} ion traps a hole strongly and a magnetically ordered state appears. 
In the ground state of the overdoped regime, on the other hand, a metallic state with Kondo-like behavior due to {\Nitwoplus} spins is realized. 
It is found that $T_{\rm N}$ decreases gradually with increasing $p_{\rm eff}$ and tends to disappear around $p_{\rm eff} = 0.18 - 0.19$ as shown in Fig. 7. 
This is reminiscent of the possible existence of QCP around $p_{\rm eff} = 0.18 - 0.19$, as formerly suggested from various experiments.~\cite{Tallon2001} 
As shown in Fig. 2, in fact, $(C_{\rm tot}-C_{\rm ph})/T$ exhibits a logarithmic increase with decreasing temperature down to the lowest temperature of 0.4 K for $(x,y)=(0.22 - 0.23, 0.05)$, which is consistent with the theoretical prediction by Moriya and Ueda in the 2D-AF quantum critical region.~\cite{Moriya} 

For $(x,y)=(0.15 - 0.18, 0.05)$ and $(0.20 - 0.23, 0.10)$, on the other hand, $(C_{\rm tot}-C_{\rm ph})/T$ increases with decreasing temperature due to the occurrence of the Kondo-like state, followed by a decrease at lower temperatures accompanied by a magnetic transition.
Moreover, {\gLT} is apparently larger than {\gN} for these samples. 
These are inconsistent with the concept of the simple QCP at which two adjacent phases in the ground state are definitely separated in the phase diagram. 
Therefore, the present results conflict with the simple QCP physics. 

It has been suggested from the magnetization,~\cite{tanabe,adachi-physc,tanabe2,tanabe3,tanabe4,tanabe5} specific heat~\cite{momono2,loram,wang} and $\mu$SR~\cite{uemura,niedermayer,bernhard} measurements that a microscopic phase separation into SC and normal-state regions takes place in a sample of the overdoped high-{\Tc} cuprates. 
Based on these results, it is possible in Ni-substituted LSCNO that a phase-separated state consisting of the magnetically ordered region observed in the underdoped regime and the metallic region observed in the overdoped regime is realized in a sample around the boundary between the magnetically ordered and metallic phases in the phase diagram. 
These explain the specific-heat results around $p_{\rm eff} = 0.10 - 0.13$ exhibiting an enhancement of $(C_{\rm tot}-C_{\rm ph})/T$ due to the occurrence of the Kondo-like state in the metallic region and a suppression of $(C_{\rm tot}-C_{\rm ph})/T$ due to the strong hole-trapping by {\Nitwoplus} in the magnetically ordered region. 
Moreover, the above explanation is also supported by larger values of $\gamma_{\rm LT}$ shown in Fig. 3 than the expected ones assuming the strong hole-trapping by {\Nitwoplus}. 
That is, as only the magnetically ordered region in a phase-separated sample is characterized by $p_{\rm eff}$, the value of $\gamma_{\rm LT}$ tends to be larger than that in the complete hole-trapping state. 
Furthermore, the distinctive behavior of $\rho$ for $(x,y)=(0.23, 0.10)$ shown in Fig. 5 is well explained taking into account the phase separation. 
That is, both the increase and following saturation of $\rho$ with a decrease of temperature below $\sim 5$ K are characteristic of a Kondo-like state in the metallic region of a sample, while the further increase of $\rho$ with decreasing temperature below 1 K is due to the strong hole-trapping by {\Nitwoplus} in the magnetically ordered region of a sample. 

The occurrence of the phase separation around the boundary between the magnetically ordered and metallic phases appears to be consistent with the results in LSCO where the superconductivity is suppressed by the application of high magnetic field~\cite{Cooper2009} or by the Zn substitution.~\cite{Risdiana2008} 
In-plane $\rho$ measurements in high magnetic fields in LSCO have revealed that a $T$-linear behavior characteristic of the quantum critical region is observed not only in a limited region of $p \sim 0.19$ but also in an extended one in the overdoped regime and coexists with the $T^2$ behavior. 
The $\mu$SR measurements for 3 \% Zn-substituted LSCZO have revealed that the Zn-induced development of the Cu-spin correlation does not vanish around $p=0.19$ as formerly indicated by Panagopoulos {\it et al}.~\cite{panago} but is weakened gradually with increasing $p$ in the overdoped regime and disappears around $x=0.30$.~\cite{Risdiana2008} 
These results are understood in terms of a phase separation into a quantum critical region with the $T$-linear resistivity and the developed Cu-spin correlation and a Fermi-liquid region with the $T^2$ resistivity. 
Accordingly, it is much convinced that the quantum phase transition in LSCO occurs not at a single point of $p$ in the ground state but in an extended region of $p$, that is, the change of the ground state upon hole doping is crossover-like due to the phase separation. 
Since the $T$-linear behavior of $\rho$ has been observed in various high-$T_{\rm c}$ cuprates~\cite{Gurvitch1987,naqib,Ando2004} and the occurrence of the phase separation has been proposed in various overdoped cuprates,~\cite{tanabe,adachi-physc,tanabe2,tanabe3,tanabe4,tanabe5,momono2,loram,wang,uemura,niedermayer,bernhard} the present results together with former ones~\cite{Cooper2009,Risdiana2008} strongly suggest that HTSC emerges around the quantum critical region affected by the phase separation.

\section{Summary}
We have carried out specific heat, $\rho$, magnetization and $\mu$SR measurements in non-SC Ni-substituted LSCNO, in order to investigate the temperature-dependent change of the electronic state and the ground state inside the pristine SC dome of LSCO without disturbing the Cu-spin correlation in the CuO$_2$ plane so much. 
In the underdoped regime, it has been found that at high temperatures there exist two kinds of hole, that is, weakly localized holes around {\Nitwoplus} forming a Zhang-Rice doublet state and itinerant holes forming a Zhang-Rice singlet state, deduced from the metallic behavior of $\rho$ and the reduced value of $S$ below 1 of Ni$^{2+}$ spins. 
With decreasing temperature, itinerant holes tend to be localized and both kinds of hole exhibit VRH conduction at low temperatures.
Finally, in the ground state, each {\Nitwoplus} ion traps a hole strongly and that a magnetically ordered state appears, evidenced from the decrease in DOS at the Fermi level in the specific heat measurements and from the appearance of a magnetic transition in the $\mu$SR measurements. 
In the overdoped regime, on the other hand, it has been found that a metallic state is realized at high temperatures, while a Kondo-like state is formed around {\Nitwoplus} spins at low temperatures, confirmed by the increase in $C_{\rm el}/T$, the logarithmic increase in $\rho$ and the negative magnetoresistance. 
It has been concluded that the ground state of non-SC LSCNO changes upon hole doping from a magnetically ordered state with the strong hole-trapping by {\Nitwoplus} to a metallic state with Kondo-like behavior due to {\Nitwoplus} spins and that the quantum phase transition is crossover-like due to the phase separation into short-range magnetically ordered and metallic regions.
Since the $T$-linear behavior of $\rho$ has been observed in various high-$T_{\rm c}$ cuprates~\cite{Gurvitch1987,naqib,Ando2004} and the occurrence of the phase separation has been proposed in various overdoped cuprates,~\cite{tanabe,adachi-physc,tanabe2,tanabe3,tanabe4,tanabe5,momono2,loram,wang,uemura,niedermayer,bernhard} it is suggested that HTSC emerges around the quantum critical region affected by the phase separation.

\section*{Acknowledgments}
We are grateful to Y. Shimizu, H. Tsuchiura, M. Ogata, T. Tohyama, K. Yamada for their helpful discussions.
We also thank A. Amato and R. Sheuermann at PSI for their technical support in the $\mu$SR measurements. 
The $\mu$SR measurements at PSI were partially supported by the KEK-MSL Inter-University Program for Oversea Muon Facilities and also by the Global COE Program "Materials Integration (International Center of Education and Research), Tohoku University", of the Ministry of Education, Culture, Sports, Science and Technology, Japan.
One of authors (Y. T.) was supported by the Japan Society for the Promotion of Science.

\end{document}